\documentclass[journal]{IEEEtran}
\usepackage{amsmath}
\usepackage{graphicx}
\usepackage{subfigure}
\usepackage{cite}
\usepackage{color}
\usepackage{epstopdf}
\usepackage{stfloats}
\usepackage{amsfonts}
\usepackage{dsfont}
\usepackage{graphicx}
\usepackage{changepage}
\begin{document}
\title{Free-Space Optical Communication With Reconfigurable Intelligent Surfaces}
\author{Liang Yang, Wang Guo, Daniel Benevides da Costa, and Mohamed-Slim Alouini
\thanks{L. Yang and W. Guo are with the Department of College of Computer Science and Electronic Engineering, Hunan University, Changsha 410082,
China, (e-mail: liangy@hnu.edu.cn, wangguo@hnu.edu.cn).}
\thanks{D. B. da Costa is with the Department of Computer Engineering, Federal University of Cear\'{a}, Sobral 62010-560, Brazil (email: danielbcosta@ieee.org).}
\thanks{M.-S. Alouini is with the CEMSE Division, King Abdullah University of Science and Technology (KAUST), Thuwal 23955-6900, Saudi Arabia (email: slim.alouini@kaust.edu.sa).}}


\maketitle

\def\baselinestretch{0.89}
%%%%%%%%%%%%%%%%%%%%%%%%%%%%%%%%%%%%%%%%%%%%%%%%%%%%%%%%%%%%%%%%%%%%%%%%%%%%%%%%%%%%%%%%%%%%%%%%%

\begin{abstract}
Despite the promising gains, free-space optical (FSO) communication is severely influenced by atmospheric turbulence and pointing error issues, which make its practical design a bit challenging. In this paper, with the aim to increase the communication coverage and improve the system performance, reconfigurable intelligent surfaces (RISs) are considered in an FSO communication setup, in which both atmospheric turbulence and pointing errors are considered. Closed-form expressions for the outage probability, average bit error rate, and channel capacity are derived assuming large number of reflecting elements at the RIS. Specifically, according to central limit theorem (CLT), while assuming multiple reflecting elements approximate expressions are proposed. It is shown that the respective accuracies increase as the number of elements at the RIS increases. Illustrative numerical examples are shown along with insightful discussions. Finally, Monte Carlo simulations are presented to verify the correctness of the analytical results.
\end{abstract}


\IEEEpeerreviewmaketitle

%%%%%%%%%%%%%%%%%%%%%%%%%%%%%%%%%%%%%%%%%%%%%%%%%%%%%%%%%%%%%%%%%%%%%%%%%%%%%%%%%%%%%%%%%%%%%%%%%%%%%%%%%%

\section{Introduction}

% [1]Optical Communication in Space: Challenges and Mitigation Techniques
% [2]Survey on free space optical communication: a communication theory perspective
% [3]Deep Learning for Improving Performance of OOK Modulation Over FSO Turbulent Channels
% [4]Effect of Pointing Errors on the Performance of Hybrid FSO/RF Networks

\IEEEPARstart{F}{ree}-space optical (FSO) communication, usually called as outdoor optical wireless communication (OWC), has been extensively investigated in the literature along the last years due to its promising gains, such as low error rates, license free operation, and high security. In addition, FSO links can be seen as a very cost effective way to provide high bandwidth links over short distances. Potential applications to FSO communication include land and space communication, such as building-to-building, satellite-to-ground, and satellite-to-satellite communications [1]. However, FSO communications are highly affected by atmospheric attenuation and turbulence of the environment which may result in performance degradation. Moreover, pointing error is another practical issue that immensely impact the system performance of FSO links [2].
A mixed FSO-radio-frequency (RF) relaying system with energy harvesting scheme was proposed in [3].

% [5]Smart Radio Environments Empowered by Reconfigurable Intelligent Surfaces:How it Works, State of Research, and Road Ahead
% [6]Wireless Communications Through Reconfigurable Intelligent surface
% [7}Wireless Communications with Reconfigurable Intelligent Surface: Path Loss Modeling and Experimental Measurement
% [8]Potential key technologies for 6G mobile communications

% [8]L. Yang, F. Meng, J. Zhang, M. O. Hasna and M. Di Renzo, "On the Performance of RIS-Assisted Dual-Hop UAV Communication Systems," in IEEE %Transactions on Vehicular Technology, doi: 10.1109/TVT.2020.3004598.
%[9]X. Hu, J. Wang, and C. Zhong, ¡°Statistical CSI based design for intelligent reflecting surface assisted MISO systems,¡± accepted to appear in %Science China: Information Science, 2020.
%[10]L. Yang, Y. Jinxia, W. Xie, M. Hasna, T. Tsiftsis and M. Di Renzo, "Secrecy Performance Analysis of RIS-Aided Wireless Communication %Systems," in IEEE Transactions on Vehicular Technology, doi: 10.1109/TVT.2020.3007521.
% [11] L. Yang, Y. Yang, M. O. Hasna and M. -S. Alouini, "Coverage, Probability of SNR Gain, and DOR Analysis of RIS-Aided Communication %Systems," in IEEE Wireless Communications Letters, vol. 9, no. 8, pp. 1268-1272, Aug. 2020, doi: 10.1109/LWC.2020.2987798.

% [12]Visible Light Communications via Intelligent Reflecting Surfaces: Metasurfaces vs Mirror arrays
% [13]Performance of Wireless Optical Communication With Reconfigurable Intelligent Surfaces and Random Obstacles
% [14]Intelligent Reconfigurable Reflecting Surfaces for Free Space Optical Communications

With the commercialization of the fifth generation (5G) wireless systems, the research community has devoted the attention to the design of  beyond 5G communications. Under this perspective, a new disruptive concept, called reconfigurable intelligent surface (RIS), has arisen as a promising technology to fulfill the envisaged future requirements. The RIS consists of man-made electromagnetic materials that can intelligently control the characteristics (e.g., electromagnetic properties) of the propagation signal, such as amplitude, phase, and polarization [4-7]. Recently, the RIS has been widely studied by virtue of its advantages and harmonious integration with wireless communication systems. For instance, the
RIS-assisted unmanned aerial vehicle (UAV) communication was studied in [8], the highly closed-form accurate approximations of the channel distribution of the RIS-assisted system were derived in [9], the RIS-aided system security was investigated in [10], and the RIS-assisted system coverage analysis was analyzed in [11].
In addition, similar to the deployment of the RIS in microwave band operations, the optical RIS has also attracted considerable attention since it can customize the reflecting incident beam, control the beam intensity, phase, frequency, and polarity, as well as adjust the orientation of the output beam due to user's movement [12]. In [13], a new pointing error model caused by beam jitter and intelligent channel reconfigurable node (ICRN) jitter was presented, in which a geometric and misalignment losses (GML) model was established to study the influence of size, position, and direction of the optical RIS on the FSO channel. However, they did not take the atmospheric turbulence into consideration.

In this paper, differently from previous works, we investigate the FSO communication with an optical RIS in the presence of both atmospheric turbulence and pointing errors. Closed-form expressions for the outage probability, average bit error rate (BER), and channel capacity are derived assuming large number of reflecting elements at the RIS. Specifically, due to the use of central limit theorem (CLT), while assuming multiple reflecting elements approximate expressions are proposed. It is shown that the approximations become tighter as the number of reflecting units at the RIS increases. Illustrative numerical examples are shown along with insightful discussions. Finally, Monte Carlo simulations are presented to verify the correctness of the analytical results. To the best of the authors' knowledge, such kind of system setup in conjunction to the proposed analysis have not been investigated in the literature yet.

%To the best of the authors' knowledge, we consider a FSO communication system based on the optical RIS in this paper.
%Suppose FSO communication is applied between two buildings. Due to non line of sight (NLOS) limitation or too
%long distance, the FSO system can not communicate or the communication performance is poor.
%Therefore, the optical RIS is placed on another building to reflect the FSO beam to improve system performance. First, the source has
%a direct link to the RIS. Then, the RIS reflects the FSO beam to the destination.
%Similar to reference [13], the single branch system and multiple branches
%system are studied, and the system performance is analyzed. But different from [13], we consider not only beam
%jitter and ICRN, but also atmospheric turbulence. Then, we derive the exacted expression of the probability density function (PDF), outage
%probability, bit error rate (BER) and channel capacity for single breach. Through the central limit theorem (CLT), we approximately analyze
%the performance of the system with multiple branches. Finally, the analysis results involving Meijer's
%G-function and Q-function are verified by Monte-Carlo method.

%%%%%%%%%%%%%%%%%%%%%%%%%%%%%%%%%%%%%%%%%%%%%%%%%%%%%%%%%%%%%%%%%%%%%%%%%%%%%%%%%%%%%%%%%%%%%%%%%%%%

\section{System and Channel Models}
We assume an RIS-aided FSO communication system composed of an optical source, an optical RIS, and a destination, as shown in Fig. 1. Due to  the obstruction of buildings, there is no direct link between the source and destination, with the communication being performed only through the RIS, which is placed strategically on a building to provide connectivity between source and destination. The RIS is formed by $N$ reflecting elements, while the source and destination are multi-aperture devices. The received signal at the destination is given by
\begin{align}
 y=\sum_{k=1}^{N}h_{k}s_{k}+n,
\end{align}
where $s_{k}$ is the signal intensity with average power $P_t$, $h_{k}=h_{a_{k}}h_{p_{k}}$ is the
cascaded channel from the source to the destination through the $k$-th reflecting element, and $n$
is the AWGN term with zero mean and variance $N_{0}$.
Let $L_{1,k}$ and $L_{2,k}$ denote the distance between source and the RIS, and
between the RIS and destination, respectively. Considering the far field case, we assume
$L_{1,k}=L_1$, $L_{2,k}=L_2$ and $L_1+L_2=L$. Based on this assumpation, all the links suffer
from the independent and identically distributed (i.i.d.) fading.

\begin{figure}
 \centering
 \includegraphics[scale=0.35]{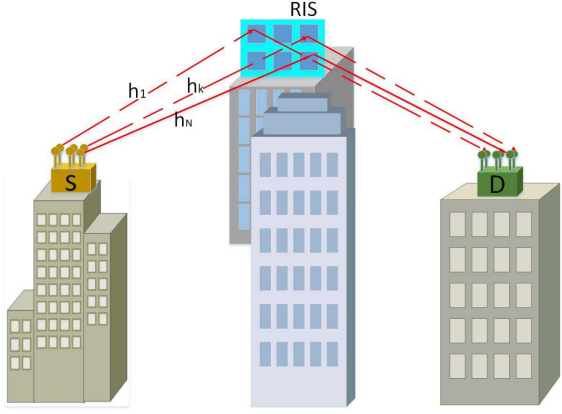}
 \caption{Block diagram of the RIS-aided FSO communication system.}
\end{figure}

\subsection{Atmospheric Turbulence}
% [15]Further results on the capacity of free-space optical channels in turbulent atmosphere
% [16]Table of Integrals, Series and Products, 7th ed
Under atmospheric turbulence, $h_{a_{k}}$ can be modeled by a Gamma-Gamma distribution which measures the moderate-to-strong atmospheric
turbulence fading [14]. Thus, the probability density function (PDF) of $h_{a_{k}}$ can be expressed as
\begin{align}
f_{h_{a}}\left ( h_{a_{k}} \right ){=}\frac{2\left ( \alpha \beta  \right )^{\frac{\alpha {+}\beta }{2}}}{\Gamma \left ( \alpha  \right )\Gamma \left ( \beta  \right )}h_{a_{k}}^{\frac{\alpha {+}\beta }{2}{-}1}K_{\alpha {-}\beta }(2\sqrt{\alpha \beta h_{a_{k}}} ),
%h_{a}{>} 0,
\end{align}
where $K_{v}(\cdot)$ represents the second kind of $v$-order modified Bessel function [15, Eq. (8.407)] and $\Gamma(\cdot)$ denotes the Gamma function [15, Eq. (8.310)]. In addition, from [14], the following parameters are determined by the atmospheric condition: $\alpha {=}\left [\rm{exp}\left ( {\frac{0.49\sigma _{R}^{2}}{\left( 1+0.18\kappa ^{2}+0.56\sigma _{R}^{\frac{12}{5}} \right)^{\frac{7}{6}}}}{-}1\right ) \right ]^{-1}$ and $\beta {=}\left [\rm{exp}\left ( \frac{0.51\sigma _{R}^{2}\left ( 1+0.69\sigma_{R}^{\frac{12}{5}}  \right )^{-\frac{5}{6}}}{\left ( 1+0.9\kappa^{2} +0.62\kappa ^{2}\sigma _{R}^{\frac{12}{5}}\right
)^{\frac{5}{6}}} \right ){-}1 \right ]^{-1}$, where $\sigma_{R}^{2}{=}0.5C_{n}^{2}k_{w}^{\frac{7}{6}}L^{\frac{11}{6}}$ and $\kappa^{2}{=}\frac{k_{w}D_{a}^{2}}{4L}$. The optical wave number can be calculated by $k_{w}{=}\frac{2\pi }{\lambda _{w}}$,
the wavelength is expressed by $\lambda _{w}$, the diameter of the receiver aperture is denoted by
$D_{a}{=}2a$, and $C_{n}^{2}$ means the index of the refractive parameter.

\subsection{Pointing Error}
Pointing error is affected by the beam jitter and ICRN jitter [12], where the beam jitter means that the beam vibrates
due to the jitter at the transmitter and ICRN jitter refers to the normal vector deflection of reflection
surface caused by the jitter of the ICRN surface. For a single intelligent optical channel, the pointing error model
has been described in [Figs. 2 and 3, 12], where the superimposed pointing error angle $\theta _{k}^{'}$ is the angle between the ICRN reflection point and the actual incident point at the receiver, and it can be modeled by a Rayleigh distribution. In addition, $\theta _{k}^{'}$ can be computed by $\theta _{k}^{'}=\sqrt{\theta _{x_{k}}^{'2}+\theta _{y_{k}}^{'2}}$ [12], where
$\theta _{x_{k}}^{'}\approx \left ( 1+\frac{L_1}{L_2} \right )\theta _{x_{k}}+2\beta _{x_{k}}$
and $\theta _{y_{k}}^{'}\approx \left ( 1+\frac{L_1}{L_2} \right )\theta _{y_{k}}+2\beta _{y_{k}}$, with
\{$\theta _{x_{k}}$, $\theta _{y_{k}}\}\sim\mathcal N\left(0,\sigma ^{2}_{\theta}\right )$ and
\{$\beta _{x_{k}}$, $\beta _{y_{k}}$\}$\sim\mathcal N\left(0,\sigma^{2} _{\beta}\right )$, where $\sim\mathcal N(\mu, \sigma ^{2})$
indicates a Normal distribution with mean $\mu$ and variance $\sigma^{2}$.
The PDF of $\theta _{k}^{'}$ can be written as
\begin{align}
f_{\theta _{k}^{'}}(\theta _{k}^{'}){=}\frac{\theta _{k}^{'}}{\left(1{+}\frac{L_1}{L_2}\right)^{2}\sigma ^{2}_{\theta}{+}4\sigma _{\beta}^{2}}e^{-\frac{\theta _{k}^{'2}}{2\left(1{+}\frac{L_1}{L_2}\right)^{2}\sigma ^{2}_{\theta}{+}8\sigma _{\beta}^{2}}}.
\end{align}
From [12], the distance $r_{k}$ from the center of the receiver to the receiving spot can be expressed as
$r_{k}={\rm{tan}}\theta _{k}^{'}L_2\approx \theta _{k}^{'}L_2$. Therefore, the channel fading caused by pointing error
can be approximated by $h_{p_{k}}\approx A_{0}{\rm{exp}}\left ( -\frac{2r_{k}^{2}}{\omega_{zeq}^{2} } \right )$, where
$A_{0}=\rm{erf}^{2} (\nu )$, $\omega_{zeq}^{2}=\omega_{z}^{2}\frac{\sqrt{\pi}{\rm{erf}}(\nu)}{2\nu e^{-\nu ^{2}} }$,
$\rm{erf}(.)$ denotes the error function [15, Eq. (8.25)], $\nu {=}\sqrt{\frac{\pi}{2}}\frac{a}{\omega _{z}}$ represents the ratio of aperture radius to beam width, $\omega_{z}$ denotes the beam width and can be computed as $\omega_{z} = \phi L$,
and $\phi $ stands for the divergence angle of the beam. Thus, the PDF of $h_{p_{k}}$ can be expressed as
\begin{align}
f_{h_{p_{k}}}\left ( h_{p_{k}} \right )=\frac{c}{A_{0}}\left ( \frac{h_{p_{k}}}{A_{0}} \right )^{c-1},\,\,\,0<h_{p_{k}}<A_{0},
\end{align}
where
\begin{align}
c=\frac{\omega _{zeq}^{2}}{4\sigma _{\theta}^{2} L^{2}+16\sigma _{\beta}^{2}L_2^{2}}.
\end{align}

% [17]BER performance of FSO links over strong atmospheric turbulence channels with pointing errors
\subsection{Statistical Distribution}
% [18] J. G. Proakis, Digital Communications, 5th ed. New York: McGrawHill, 2008.

As mentioned in [12], it is assumed that the transmitter transmits signals to $N$ RISs, and
each RIS directly reflects the signals to the receiver. All beams are centered on the receiver, and the
receiver receives all the energy of the beam.
Thus, the instantaneous SNR can be formulated as
\begin{align}
\gamma=\sum_{k=1}^{N}\gamma_{k}=\bar{\gamma }\sum_{k=1}^{N}h_{k}^{2}=\bar{\gamma }Z,
\end{align}
where $\gamma_{k}$ means the instantaneous SNR of the $k$-th channel and $\bar{\gamma}=\frac{P_{t}}{N_{0}}$ denotes the average SNR..

From [14], the PDF of $B=h_{k}^{2}$ can be obtained as
\begin{align}
f_{B}\left ( x  \right ){=}& \frac{\alpha \beta c }{2\sqrt{x}\Gamma \left ( \alpha  \right )\Gamma  ( \beta  )A_{0}}\, \nonumber\\
&\times{\rm{G}}_{1,3}^{3,0}
\left [{{\frac{\alpha \beta\sqrt{x} }{A_{0}}}\left |{ \begin{matrix} {c}
\\ {c{-}1,\alpha{-}1,\beta{-}1} \\ \end{matrix} }\right . }\right ]
,
\end{align}
where $G_{.,.}^{.,.}(.)$ denotes the Meijer's G-function [16, Eq. (07.34.02.0001.01)]. Thus, the mean and variance of $B$ can be expressed as
\begin{align}
m_{1} {=} {\rm{E}}[B]{= }\frac{c A_{0}^{2}\Gamma \left ( c{+}2 \right )\Gamma \left ( \alpha {+}2 \right )\Gamma \left ( \beta {+}2 \right )}{\alpha ^{2}\beta ^{2}\Gamma \left ( \alpha  \right )\Gamma \left ( \beta \right )\Gamma \left ( c{+}3 \right )},
\end{align}
\begin{align}
\delta_{1}^{2} {=} {\rm{VAR}}[B] {=} \frac{cA_{0}^{4}\Gamma \left ( c{+}4 \right )\Gamma \left ( \alpha {+}4 \right )\Gamma \left ( \beta{ +}4 \right )}{\alpha ^{4}\beta ^{4}\Gamma \left ( \alpha  \right )\Gamma \left ( \beta \right )\Gamma \left ( c{+}5 \right )}{-}m_{1}^{2}.
\end{align}
From (7), the PDF of $\gamma_{k}$ can be written as
\begin{align}
f_{\gamma_{k} }\left ( \gamma  \right ){=}& \frac{\alpha \beta c}{2\sqrt{\bar{\gamma }}\sqrt{\gamma }\Gamma \left ( \alpha  \right )\Gamma  ( \beta  )A_{0}}\nonumber\\
&\times{\rm{G}}_{1,3}^{3,0}
\left [{{\frac{\alpha \beta }{A_{0}}\sqrt{\frac{\gamma }{\bar{\gamma }}}}\left |{ \begin{matrix} {c}
\\ {c{-}1,\alpha{-}1,\beta{-}1} \\ \end{matrix} }\right . }\right ].
\end{align}

Using (10) to evaluate the system performance is very difficult. On the basis of the CLT [17] and assuming large $N$, $Z$ can be well-approximated by a Gaussian random variable with mean value equals to $m = m_{1}\times N$ and variance given by $\delta^{2} = \delta_{1}^{2}\times N$. Thus, the PDF of $\gamma$ can be written as
\begin{align}
f_{\gamma} \left ( x \right )\approx \frac{1}{\sqrt{2\pi \delta ^{2}}\bar{\gamma }}e^{-  \frac{\left (x-\bar{\gamma }m\right )^{2}}{2\bar{\gamma }^{2}\delta ^{2}}}.
\end{align}

Relying on [15, Eq. (3.322.2)] and performing the appropriate substitutions, the moment generating function (MGF), which is defined by
$M_{\gamma }\left ( s \right )={\rm{E}}\left[{{e}}^{-s\gamma } \right ]$, with $\rm{E}[\cdot]$ denoting expectation, can be derived as
\begin{align}
M_{\gamma }\left ( s \right )\approx\frac{1}{2}e^{\frac{\bar{\gamma }^{2}\delta ^{2}s}{2}-\frac{m^{3}}{2\delta ^{2}}-\frac{\bar{\gamma }m}{2}}\left [ 1-\Phi \left ( \frac{s\bar{\gamma }\delta }{\sqrt{2}} -\frac{m}{2\delta }\right ) \right ],
\end{align}
where $\Phi(\cdot)={\rm erf}(\cdot)$.

In addition, from (11) and [15, Eq. (3.462.1)], the generalized moments of $\gamma$, which is given by ${\rm{E}}\left [ \gamma ^{n} \right ]=\int_{0}^{\infty }\gamma ^{n}f_{\gamma }\left ( \gamma  \right )d\gamma$, can be derived as
\begin{align}
{\rm{E}}\left [ \gamma ^{n} \right ]\approx\frac{\bar{\gamma}^{n}\delta ^{n}}{\sqrt{2\pi }}e^{\frac{-m^{2}}{4\delta ^{2}}}\Gamma \left ( n+1 \right )D_{-n-1}\left ( -\frac{m}{\delta } \right ),
\end{align}
where $D_{v}(.)$ is the parabolic cylinder functions.

Finally, from (13), the $n$-order amount of fading (AF), which is defined as $AF_{\gamma }^{(n)}=\frac{{\rm{E}}\left [ \gamma ^{n} \right ]}{{\rm{E}}\left [ \gamma  \right ]^{n}}-1$ and arises as an important index to measure the system performance, can be obtained as
\begin{align}
AF_{\gamma }^{(n)}\approx\frac{\delta ^{n}}{\sqrt{2\pi }m^{n}}\Gamma \left ( n+1 \right )e^{-\frac{m^{2}}{4\delta ^{2}}}D_{-n-1}\left ( -\frac{m}{8} \right )-1.
\end{align}

%%%%%%%%%%%%%%%%%%%%%%%%%%%%%%%%%%%%%%%%%%%%%%%%%%%%%%%%%%%%%%%%%%%%%%%%%%%%%%%%%%%%%%%%%%%%%%%%%%%%%%%
\section{Performance Analysis}
%Outage probability, average BER, and channel capacity are important indexes to measure the performance of
%communication systems. In this section, we carry out a performance analysis of the proposed system setup and derive closed-form expressions for the aforementioned metrics.

% [19]Moment Generating Function of the Generalized ¦Á?¦Ì Distribution with Applications
% [20]Capacity of Fading Channels with Channel Side Information
% [21]Wolfram, The Wolfram functions site, Available: http://functions.wolfram.com.
In this section, the outage probability, BER, and channel capacity are derived as important indexes to measure the system performance.

\subsection{Outage Probability}
Outage probability can be defined as the probability that the instantaneous SNR falls below a given threshold $\gamma_{th}$, being mathematically formulated as
\begin{align}
P_{out}=\Pr\left ( \gamma \leq  \gamma_{th} \right )=\int_{0}^{\gamma _{th}}f_{\gamma }\left ( \gamma  \right )d\gamma .
\end{align}
From [15, Eqs. (2.322.1) and (2.322.2)], one can get the outage probability by integrating (15) as
\begin{align}
P_{out}\approx\frac{1}{2}\left [ \Phi \left ( \frac{\gamma _{th}-m\bar{\gamma }}{\sqrt{2}\bar{\gamma }\delta }\right ) -\Phi \left ( -\frac{m}{\sqrt{2}\delta } \right )\right ].
\end{align}

In order to get further insights to the outage performance, asymptotic outage analysis is now carried.
By setting $\bar{\gamma}\rightarrow \infty $, (10) can be written as [16, Eq.(07.34.06.0006.01)]
\begin{align}
f_{\gamma_{k}}(\gamma) \rightarrow \sum_{i=1}^{3}\frac{\prod_{j=1,j\neq i}^{3}\Gamma \left ( b_{j}-b_{i} \right )}{\Gamma \left ( c-b_{i} \right )}\left ( \frac{\alpha \beta }{A0} \sqrt{\frac{\gamma }{\bar{\gamma }}}\right )^{b_{i}},
\end{align}
where $b_{i}=(c-1, \alpha-1, \beta-1)$.

Similar [3], $f_{\gamma}(\gamma)$ is determined by the minimum value of $\varrho = {\rm{min}}(b_{i})$. Then, similar to (12), we can obtain the MGF of $\gamma_{k}$
\begin{align}
M_{\gamma_{k}}(s) \rightarrow \frac{\epsilon c\alpha ^{1+\varrho }\beta ^{1+\varrho } \Gamma \left ( \frac{1+\varrho}{2} \right )}{2\Gamma \left ( \alpha  \right )\Gamma \left ( \beta  \right )A_0^{1+\varrho} (\bar{\gamma }s)^{\frac{1+\varrho}{2}}},
\end{align}
where $\epsilon$ is the coefficient of the $\left ( \frac{\alpha \beta }{A0} \sqrt{\frac{\gamma }{\bar{\gamma }}}\right )^{\varrho}$.
According to the relationship of $M_{\gamma}(s)=(M_{\gamma_{k}}(s))^N$ [12, Eq.(33)],
 $f_{\gamma}(\gamma) = \int_{0}^{\infty }e^{s\gamma }M_{\gamma }(s)ds$ and (15),
the outage probability can be asymptotically written as
\begin{align}
P_{out} &\rightarrow  \left (  \frac{\epsilon c\alpha^{1+\varrho }  \beta^{1+\varrho}\Gamma \left ( \frac{1+\varrho}{2} \right )  }{2\Gamma \left ( \alpha  \right )\Gamma \left ( \beta  \right )A_0^{1+\varrho} \bar{\gamma }^{\frac{1+\varrho}{2}}} \right )^{N}\\ \nonumber
&\times \frac{\gamma _{th}^{\frac{(1+\varrho)N}{2}}}{\Gamma \left ( \frac{(1+\varrho)N}{2} \right )\frac{(1+\varrho)N}{2} },
\end{align}
which indicates that the diversity order equals to $\frac{(1+\varrho)N}{2}$. Therefore, increasing $N$ can improve the system performance.
\subsection{Average BER}
% [22]New Exponential Bounds and Approximations for the Computation of Error Probability in Fading Channels
For most binary correlated modulation schemes, the average BER can be calculated by the following expression [17]
\begin{align}
Pe= &\int_{0}^{\infty }Q\left ( \sqrt{2\psi \gamma } \right )f_{\gamma }\left ( \gamma  \right )d\gamma,
\end{align}
where $Q(\cdot)$ denotes the Q-function [15, Eq. (6.287.3)], and $\psi$ represents a coefficient applied to different modulation methods. For example, $\psi= 1$ for binary phase shift keying (BPSK), $\psi= 0.5$ for coherent detection of binary
frequency keying (BFSK), and $\psi= 0.75$ for coherent detection of minimum shift keying [18].

From [19, Eq.(14)] and [15, Eq.(6.287)], the Q-function can be approximately expressed as
$Q\left ( x\right )=\frac{1}{2}{\rm{erfc}}\left ( \frac{x}{\sqrt{2}} \right )\simeq \frac{1}{12}e^{-\frac{x^{2}}{2}}+\frac{1}{4}e^{-\frac{2x^{2}}{3}}.$
Considering the BPSK modulation scheme and making use of [15, Eqs. (2.322.1) and (2.322.2)], we obtain
\begin{align}
Pe\approx & \frac{1}{12}\frac{\sqrt{\pi \rho }e^{-\frac{m^{2}}{2\delta ^{2}}}}{\sqrt{2\pi \delta ^{2}}\bar{\gamma }}
e^{\rho  \varepsilon^{2}}\left ( 1-\Phi \left ( \varepsilon\sqrt{\rho } \right ) \right )\nonumber\\
&+\frac{1}{4}\frac{\sqrt{\pi \rho }e^{-\frac{m^{2}}{2\delta ^{2}}}}{\sqrt{2\pi \delta ^{2}}\bar{\gamma }}e^{\rho  \iota^{2}}\left ( 1-\Phi \left ( \iota\sqrt{\rho } \right ) \right ),
\end{align}
where $\varepsilon =1-\frac{m}{\bar{\gamma }\delta ^{2}}$, $\iota=\frac{4}{3}-\frac{m}{\bar{\gamma }\delta ^{2}}$, $\rho =\frac{\bar{\gamma }^{2}\delta ^{2}}{2}$.

\subsection{Channel Capacity}

% [23]Novel Unified Expressions for Error Rates and Ergodic Channel Capacity Analysis over Generalized Fading Subject to AWGGN
For the channel capacity, it can be calculated as [20]
\begin{align}
C = \int_{0}^{\infty }{\rm{log}}_{2}\left ( 1+x \right )f_{\gamma }\left ( x \right )dx.
\end{align}
From [21], it is shown that $\log_{2}\left ( 1+x \right )$ can be well-approximated by
$\log_{2}\left ( 1+x \right )\approx \sum_{i=1}^{4}\eta_{i} e^{-\zeta _{i}x}$, where $\eta =\left [ 9.331,-2.635,-4.032,-2.388 \right ]$
and $\zeta = \left [ 0.000, 0.037, 0.004, 0.274 \right ]$. Then, the channel capacity can be derived as
\begin{align}
C &\approx\frac{1}{2}e^{-\frac{m^{2}}{2\delta ^{2}}}\sum_{i=1}^{4}\eta _{i}e^{\frac{\bar{\gamma }^{2}\delta ^{2}}{2}\left ( \zeta _{i} -\frac{m}{\bar{\gamma }\delta ^{2}}\right )^{2}}\nonumber\\
&\times\left [ 1-\Phi \left ( \frac{\zeta _{i}\bar{\gamma }\delta }{\sqrt{2}}-\frac{m}{\sqrt{2}\delta } \right ) \right ].
\end{align}

%%%%%%%%%%%%%%%%%%%%%%%%%%%%%%%%%%%%%%%%%%%%%%%%%%%%%%%%%%%%%%%%%%
\section{Numerical Results and Discussions}
In this section, representative numerical results are provided to illustrate the performance of the RIS-aided FSO
communication system.
The impact of the key system parameters on the system performance is investigated.
Monte Carlo simulations are also presented to corroborate the proposed analysis and derived expressions.
Unless otherwise stated, we have considered in the plots the following parameters: $\sigma_{\theta}=1$ mrad, $\sigma_{\beta}=0.5$ mrad, $w_z=120$ cm, $a=10$ cm, $\alpha=15$, $\beta=10$, $L_1=L_2=150$ m [12]. In addition, when a single path (direct link, without RIS) is considered, we set the path distance as $L=100$ m and the standard deviation of pointing error angle as $\sigma_{\theta}=1$ mrad.

\begin{figure}
 \centering
 \includegraphics[scale=0.55]{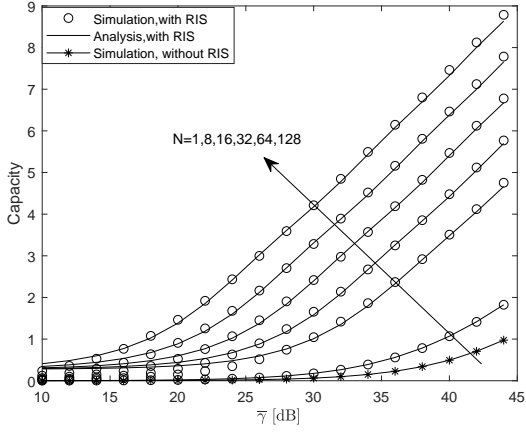}
 \caption{Channel capacity of the RIS-aided FSO system for different values of $N$.}
\end{figure}

\begin{figure}
 \centering
 \includegraphics[scale=0.55]{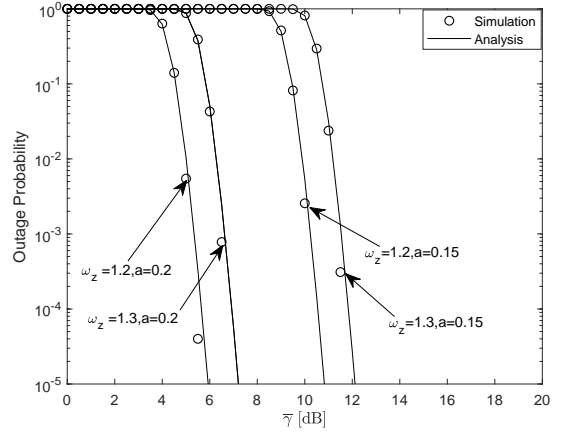}
 \caption{Outage probability of the RIS-aided FSO systems for different values of $\omega_{z}$ and $a$.}
\end{figure}

\begin{figure}
 \centering
 \includegraphics[scale=0.55]{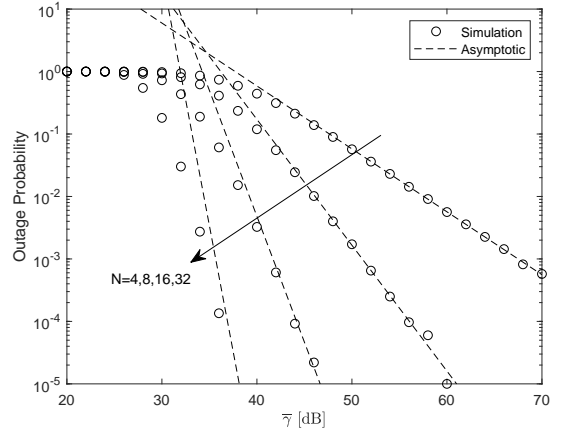}
 \caption{Outage probability of the RIS-aided FSO system for different values of $N$ ($\alpha = 6.5$, $\beta = 6.0$, $\sigma_{\theta}=20$ mrad, $\sigma_{\beta}=20$ mrad, and $c = 0.5$).}
\end{figure}

\begin{figure}
 \centering
 \includegraphics[scale=0.55]{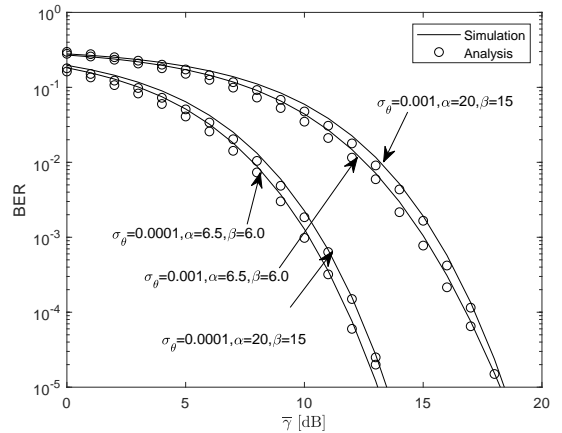}
 \caption{BER of the RIS-aided FSO system for different values of $\alpha$, $\beta$, and $\sigma_{\theta}$.}
\end{figure}

In Fig. 2, the channel capacity versus $\bar{\gamma}$ is plotted for different values of $N$. Note that although the path distance when $N=1$ is $300$ m, its capacity performance slightly improves under high SNR regime when compared with the direct link assuming $100$ m as path distance. Observe also that the capacity performance significantly improves as $N$ increases.
%In addition, as expected, due to the use of CLT approach, the analytical curves do not coincide with the simulation ones at high SNR when $N$ is %small. However, the approximations become very tight for large values of $N$ (e.g., $N=128$).

Fig. 3 plot the outage probability, versus $\bar{\gamma}$ for different values of $\omega_{z}$ and $a$. We set $N=128$ and the other parameters are consistent with those of Fig. 2. Similar conclusions to those ones attained in Fig. 2 are attained. In addition, it can be observed that the outage performance can be improved by decreasing the value of $\frac{w_{z}}{a}$. Fig. 4 plots the asymptotic outage curves, which shows that the diversity order increases with $N$, corroborating our analysis.

%
%BPSK modulation technology is applied in this system. It can be seen from Fig. 3 that the BER
%performance of both single branch system and multiple branches system has been improved compared
%with direct path. By using the CLT, the $N$ required for BER simulation results to coincide with the
%analysis results is smaller than that for outage probability.
%

%Fig. 4 shows the channel capacity performance, and the parameters are still set as shown in Figs. 2 and 3.
%It is clear that applying RISs can markedly improve the channel capacity performance.
%And with the increase of reflection units, better channel capacity can be obtained.
Finally, the effect of different system parameters (e.g.,  $\alpha$, $\beta$, and $\sigma_{\theta}$) on the BER
performance is studied in Fig. 5. We set $L_1 = 350$ m, $L_2 = 250$ m, $N = 128$, $a=20$ cm. It can
be clearly seen that the BER probability under strong turbulence is better than that of under moderate turbulence.
In addition, with the change of atmospheric turbulence, the BER performance is slightly affected. One can also see that $\sigma_{\theta}$ has a great impact in the performance, i.e., when the value of $\sigma_{\theta}$ increases, the BER becomes worse.

%%%%%%%%%%%%%%%%%%%%%%%%%%%%%%%%%%%%%%%%%%%%%%%%%%%%%%%%%%%%%%%%%%%%%%%%%%%%%%%%%%%%%%%%
\section{Conclusion}
In this paper, a FSO system aided by RISs was investigated. Closed-form expressions for the outage probability, average BER, and channel capacity were derived. It was shown that the deployment of RISs can significantly improve the performance of FSO communication systems. In addition, our results reveal that pointing error has a great impact on the system performance when compared with the atmospheric turbulence.


\begin{thebibliography}{1}
\bibitem{1}
H. Kaushal and G. Kaddoum, ``Optical communication in space: challenges and mitigation techniques," \emph{IEEE Commun. Surveys  Tuts.}, vol. 19, no. 1, pp. 57-96, Firstquarter 2017.
\bibitem{2}
M. A. Khalighi and M. Uysal, ``Survey on free space optical communication: a communication theory perspective," \emph{IEEE Commun. Surveys  Tuts.}, vol. 16, no. 4, pp. 2231-2258, Fourthquarter 2014.
\bibitem{3}
J. Chen et al., ``A novel energy harvesting scheme for mixed FSO-RF relaying systems," \emph{IEEE Trans. Veh. Technol.}, vol. 68, no. 8, pp. 8259-8263, Aug. 2019.
\bibitem{4}
M. Di Renzo et al., "Smart radio environments empowered by reconfigurable intelligent surfaces: how it works, state of research, and road ahead," \emph{IEEE J. Sel. Areas Commun.}, DOI: 10.1109/JSAC.2020.3007211.
\bibitem{5}
E. Basar, M. Di Renzo, J. De Rosny, M. Debbah, M. Alouini, and R. Zhang, ``Wireless communications through reconfigurable intelligent surfaces," \emph{IEEE Access}, vol. 7, pp. 116753-116773, 2019.
\bibitem{6}
W. Tang, M. Chen, X. Chen, J. Dai, Y. Han, M. Renzo, Y. Zeng, S. Jin, Q. Cheng, T. Cui, ``Wireless communications with reconfigurable intelligent surface: path loss modeling and experimental measurement," [Online]. Available:https://arxiv.org/abs/1911.05326.
\bibitem{7}
Y. Yuan, Y. Zhao, B. Zong, et al. ``Potential key technologies for 6G mobile communications," \emph{Sci. China Inf. Sci.}, 63, 183301 (2020).
\bibitem{8}
L. Yang, F. Meng, J. Zhang, M. O. Hasna, and M. Di Renzo, ``On the performance of RIS-assisted dual-hop UAV communication systems," \emph{IEEE Trans. Veh. Technol.}, DOI: 10.1109/TVT.2020.3004598.
\bibitem{9}
L. Yang, F. Meng, Q. Wu, D. B. da Costa, and M. Alouini, ``Accurate closed-form approximations to channel distributions of RIS-aided wireless systems," \emph{IEEE Wireless Commun. Let.}, DOI: 10.1109/LWC.2020.3010512.
\bibitem{10}
L. Yang, J. Yang, W. Xie, M. O. Hasna, T. Tsiftsis, and M. Di Renzo, ``Secrecy performance analysis of RIS-aided wireless communication systems," \emph{IEEE Trans. Veh. Technol.}, DOI: 10.1109/TVT.2020.3007521.
\bibitem{11}
L. Yang, Y. Yang, M. O. Hasna, and M. Alouini, ``Coverage, probability of SNR gain, and DOR analysis of RIS-aided communication systems," \emph{IEEE Wireless Commun. Let.}, vol. 9, no. 8, pp. 1268-1272, Aug. 2020.
\bibitem{12}
H. Wang, Z. Zhang, B. Zhu, J. Dang, L. Wu, L. Wang, K. Zhang, Y. Zhang, ``Performance of wireless optical communication with reconfigurable intelligent surfaces and random obstacles," [Online]. Available:https://arxiv.org/abs/2001.05715.
\bibitem{13}
M. Najafi, B. Schmauss, R. Schobe, ``Intelligent reconfigurable reflecting surfaces for free space optical communication," [Online]. Available:https://arxiv.org/abs/2005.04499.
\bibitem{14}
W. Gappmair,  ``Further results on the capacity of free-space optical channels in turbulent atmosphere," \emph{IET Commun.}, vol. 5, no. 9, pp. 1262-1267, Jun. 2011.
\bibitem{15}
I. S. Gradshteyn and I. M. Ryzhik, \emph{Table of integrals, series and products,} 7th ed. San Diego, CA, USA: Academic, 2007.
\bibitem{17}
Wolfram, The Wolfram functions site, Available: http://functions.wolfram.com.
\bibitem{18}
J. G. Proakis, Digital Communications, 5th ed. New York: McGrawHill, 2008.
\bibitem{19}
A. M. Magableh and M. M. Matalgah, ``Moment generating function of the generalized $\alpha$-$\mu$ distribution with applications," \emph{IEEE Commun. Let.}, vol. 13, no. 6, pp. 411-413, Jun. 2009.
\bibitem{20}
M. Chiani, D. Dardari, and M. K. Simon, ``New exponential bounds and approximations for the computation of error probability in fading channels," \emph{IEEE Trans. Wireless Commun.}, vol. 2, no. 4, pp. 840-845, Jul. 2003.
\bibitem{21}
A. J. Goldsmith and P. P. Varaiya, ``Capacity of fading channels with channel side information," \emph{IEEE Trans. Inf. Theory}, vol. 43, no. 6, pp. 1986-1992, Nov. 1997.
\bibitem{22}
E. Salahat and A. Hakam, ``Novel unified expressions for error rates and ergodic channel capacity analysis over generalized fading subject to AWGGN," \emph{2014 IEEE GCC}, Austin, TX, 2014, pp. 3976-3982.

\end{thebibliography}
\end{document}